\documentclass[11pt]{article}

\usepackage[a4paper,margin=1in]{geometry}
\usepackage[T1]{fontenc}
\usepackage{lmodern}

\usepackage{amsmath,amssymb,amsthm}
\usepackage{mathtools}

\usepackage{enumitem}

\usepackage{tikz}
\usetikzlibrary{arrows.meta,positioning,shapes.geometric}

\usepackage{cite}
\usepackage{float}
\usepackage{hyperref}

\newcommand{\ket}[1]{\lvert #1\rangle}
\newcommand{\bra}[1]{\langle #1\rvert}
\newcommand{\braket}[2]{\langle #1\mid #2\rangle}

\newtheorem{prop}{Proposition}
\newtheorem{remark}{Remark}

\title{What Kind of World Supports Darwinian Evolution?\\Quantum Foundational Options}
\author{Partha Ghose\thanks{Email: partha.ghose@gmail.com}\\
Tagore Centre for Natural Sciences and Philosophy,\\
Rabindra Tirtha, New Town, Kolkata 700156, India}
\date{}

\begin{document}
\maketitle

\begin{abstract}
Darwinian evolution requires (i) heritable records, (ii) repeatable copying with variation, and (iii) routine irreversibility.
Categorical quantum mechanics (CQM) makes precise why ``copy'' and ``delete'' are not generic quantum operations: they exist only for a realized \emph{classical data} sector (a preferred basis/observable; a commutative structure).
Decoherence explains how a pointer basis can be selected dynamically, but it does not by itself select a unique outcome.
This motivates a neutral presentation of the main ontological options (unique-history, decohered multiplicity, agent-relative facticity, and a stochastic foundation with variable diffusion). We also note the relevance of the ``agency constraint'' argued by
Adlam--McQueen--Waegell: in a strictly coherent, basis-unselected ``purely quantum'' regime, minimal agency fails due to no-cloning and linearity, which sharpens the role of classical resources for record-based processes. Extended Wigner's Friend scenarios then serve as a stress test, since they treat ``friends'' simultaneously as coherent quantum systems and as agents possessing stable records. Finally, a stochastic-mechanics foundation (with variable diffusion) offers a continuous bridge between quantum and classical regimes, and suggests a principled way to implement measurement update as conditioning plus a time-symmetric minimal-change rule.
\end{abstract}

\section{Introduction}
Darwinian evolution is commonly formulated as a physically instantiated cycle of \emph{records} (genotypes), \emph{copying with occasional variation}, and \emph{selection} sustained by irreversible dissipation. From a foundational standpoint this raises a sharp question: what becomes of such a record-driven process if the world is \emph{fundamentally quantum mechanical}, with no preferred basis and only unitary dynamics? At face value, copying and deleting are constrained by the no-cloning \cite{WZ1982} and no-deleting \cite{PBNoDelete2000} principles, and the distinction between ``a record'' and ``a coherent superposition'' is basis-dependent. In quantum theory, universal deletion of an unknown state by a reversible operation is impossible (no-deleting), which mirrors the no-cloning obstruction; physically, ``deletion'' is implemented as irreversible discarding into an environment.
 
The first aim of this note is to isolate the minimal structural ingredients required for evolution to be even \emph{well-posed} as a physical process in a quantum mechanical world. Categorical quantum mechanics provides a clean language for this: ``copy'' and ``delete'' are not generic quantum operations, but arise only on a realized \emph{classical data} sector (a commutative structure selecting a preferred observable/basis) \cite{CK2017}. Decoherence and environment-induced superselection explain how such a pointer basis can emerge dynamically and why macroscopic records are stable \cite{Zurek2003}, while leaving open the distinct question of whether outcomes are unique in a single history.

The second aim is to present, in neutral fashion, the main ontological packages that locate facts and histories in different ways (unique-history, unitary-plus-decohered multiplicity, and agent-relative approaches), and to indicate how extended Wigner's Friend setups stress-test any account that treats embedded observers both as coherent quantum systems and as record-bearing agents \cite{FR2018,Brukner2018,Proietti2019}. In this connection we note the ``agency constraint'' argued by Adlam--McQueen--Waegell \cite{Adlam2025}: a strictly coherent, basis-unselected ``purely quantum'' regime cannot sustain minimal agency, sharpening the role of classical resources in record-based processes. Finally, we briefly indicate how a stochastic-mechanics foundation \cite{Nelson1985} with variable diffusion can offer a continuous bridge between quantum and classical regimes, and how measurement update can be implemented as conditioning together with a time-symmetric minimal-change principle \cite{Pavon1999}.

\section{Minimal physical requirements for Darwinian evolution}
A world that supports Darwinian evolution (in the ordinary sense) must supply:
\begin{enumerate}[label=(E\arabic*),itemsep=2pt]
\item \textbf{Stable records:} persistent, distinguishable states functioning as an ``alphabet'' (genotypes, memory traces).
\item \textbf{Copying with variation:} approximate replication of those records with occasional local errors.
\item \textbf{Irreversibility/deletion:} systematic loss of alternatives and disposal of records (entropy export, waste heat), so that selection is not merely reversible reshuffling.
\end{enumerate}
These are not merely biological requirements; they are structural requirements on the underlying physics: there must be a
\emph{copyable record sector} and an \emph{arrow of irreversibility} for evolution to be well-defined and effective.

\section{Categorical Quantum Mechanics: why ``classical data inside QM'' is special}
In categorical quantum mechanics, a key distinction is between generic quantum states and the special states that can serve as
\emph{classical data}. Operationally, ``classical data'' means \emph{perfectly distinguishable states relative to a chosen observable}.

Fix an orthonormal basis $\{\ket{i}\}$ (equivalently: fix an observable whose eigenstates are $\{\ket{i}\}$). Then one can define
\begin{equation}
\Delta:\ \ket{i}\longmapsto \ket{i}\otimes\ket{i}
\qquad\text{and}\qquad
\varepsilon:\ \ket{i}\longmapsto 1,
\end{equation}
i.e.\ copying and deleting \emph{basis symbols}. In CQM this is packaged as a commutative special $\dagger$-Frobenius algebra (a ``classical structure'') \cite{CK2017}.

However, $\Delta$ is not a basis-free copier. For a superposition,
\begin{equation}
\Delta(\alpha\ket{0}+\beta\ket{1})=\alpha\ket{00}+\beta\ket{11}\neq(\alpha\ket{0}+\beta\ket{1})^{\otimes 2}.
\end{equation}
Thus copying exists only \emph{relative to} a distinguished classical sector (a preferred basis/record structure).

\begin{prop}[Evolution needs a realized classical sector]
If no physically selected classical data sector exists at the relevant scales (no stable, distinguishable record basis supporting copying/deleting), then heredity and ``copying the same symbol again'' are not well-defined operations, and Darwinian evolution (as ordinarily formulated) cannot get traction.
\end{prop}

\section{Decoherence: basis-selection without outcome-selection}
Decoherence addresses (part of) the \emph{record question} by explaining why certain states become robust ``pointer states'' \cite{Zurek2003} under
interaction with an environment. Schematically, a measurement-like interaction produces correlations
\begin{equation}
\sum_i c_i\,\ket{i}_S\ket{E_0}\ \longrightarrow\ \sum_i c_i\,\ket{i}_S\ket{E_i},
\qquad \braket{E_j}{E_i}\approx \delta_{ij}.
\end{equation}
Tracing out the environment yields a reduced state approximately diagonal in the pointer basis:
\begin{equation}
\rho_S \approx \sum_i |c_i|^2\,\ket{i}\bra{i}.
\end{equation}
This is \emph{basis-selective}: diagonality is tied to the pointer basis selected by the dynamics. Re-expressing $\rho_S$ in another basis (or applying a unitary to $S$) can of course produce off-diagonal matrix elements. However, this does not by itself recover interference between the decohered pointer alternatives: the relevant phase information is stored in system--environment correlations, and genuine recoherence requires reversing those correlations (e.g.\ by acting jointly on $S$ and the pertinent environmental degrees of freedom).

Decoherence, however, does not by itself deliver a \emph{unique} outcome for a single run: the global state remains entangled.
Hence decoherence is not, by itself, a substitute for ``collapse'' if one demands a single-history ontology.

\section{Agency constraint and its relevance to records}
Adlam--McQueen--Waegell argue that \emph{agency cannot be a purely quantum phenomenon} if ``purely quantum'' is taken to mean: coherent unitary evolution without decoherence or collapse and without an emergent preferred basis \cite{Adlam2025}. Their analysis isolates minimal conditions for agency (world-model construction, deliberation over alternatives using that model, and reliable action selection) and connects the failure of those conditions to no-cloning and linearity.

\begin{remark}[Why this matters here]
Darwinian evolution is not deliberative agency, but it is still a record-driven copy-and-filter process. The agency analysis can be read as a sharpened ``no-copy/no-record'' obstruction: in a regime with no preferred copyable record basis, not only agency but any stable information-processing architecture is threatened, and with it the physical preconditions for heredity.
\end{remark}

\section{Three Logically Distinct Questions and the Options}
To keep the discussion transparent, it is helpful to separate three logically distinct questions:

\begin{enumerate}[label=(Q\arabic*),itemsep=2pt]
\item \textbf{Record question (classical data):} What physical mechanism (or postulate) guarantees the existence of a robust,
copyable \emph{record sector} (a preferred basis/observable) on which copying/deleting are meaningful?

\item \textbf{Outcome question (single history):} Do measurement-like interactions yield \emph{one} actual outcome in a single
macroscopic history, or do they yield only an effectively classical reduced description (or a family of decohered histories)
without selecting one history as uniquely actual?

\item \textbf{Agency/evolution question:} Once records exist, what resources make copying, discarding, and selection
(or more generally, agent-like inference and control) physically well-defined?
\end{enumerate}

CQM answers (Q1) \emph{structurally}: copying/deleting are not generic quantum operations, but exist only relative to a realized \emph{classical structure} (a commutative sector). Decoherence contributes to (Q1) by explaining how a pointer basis can be dynamically selected, but it does not by itself settle (Q2). Adlam--McQueen--Waegell sharpen (Q3): in a strictly coherent regime with no preferred basis (and hence no copying), minimal agency fails, highlighting the importance of classical resources for record-based processes.

With these questions separated, one can present the main ontological packages as \emph{options}:

\subsection*{Option A: Unique-history realism (single actual macroscopic history)}
(Q2) is answered affirmatively: there is a single realized history with definite outcomes, supplied either by objective-collapse
dynamics \cite{GRW1986,Pearle1989,GPR1990,Diosi1989,Penrose1996,BassiGhirardi2003,BassiEtal2013} or by an added-variable/trajectory ontology. A standard example of the added-variable route is de~Broglie--Bohm (Bohmian) mechanics, in which particle configurations (guided trajectories) are ontological and a single macroscopic history is selected without modifying the unitary Schr\"odinger evolution \cite{Bohm1952a,Bohm1952b,DGZ2013,Holland}. By contrast, Nelson's stochastic mechanics takes the fundamental ontology to be classical diffusion-like stochastic trajectories, with the Schr\"odinger equation and Hilbert-space machinery arising as an emergent ensemble-level representation rather than as primitive postulates \cite{Nelson1985}. (Q1) is then grounded by that additional structure: records are definite because outcomes are definite (or because beables/trajectories ground definiteness).

\subsection*{Option B: Unitary-only + decohered multiplicity (many effectively classical histories)}
(Q2) is not answered at the fundamental level: the universal state contains many mutually decohered record-histories \cite{Griffiths1984,GMH1990,GMH1993}.
(Q1) is answered dynamically: a record basis (pointer states) emerges and supports stable classical data within each decohered sector.

\subsection*{Option C: Agent-relative facticity (relational/QBist family)}
(Q2) is treated as agent- or interaction-indexed: outcomes are facts relative to agents/relations rather than absolute global facts \cite{Rovelli1996,CFS2002,Fuchs2010,FS2013,FMS2014}. (Q1) is tied to what is stably communicable and operationally accessible between agents, rather than to a single global preferred basis.

\subsection*{Option D: Stochastic foundation with a continuum of regimes (variable diffusion)}
On a Nelson-type stochastic-mechanics view, the fundamental ontology is a family of \emph{classical stochastic trajectories} $x(t,\omega)$ (diffusions) together with their ensemble statistics; the Hilbert-space formalism is not taken as primitive but as an \emph{effective representation} of the induced probability flow and phase-like variables in suitable regimes \cite{Nelson1985}. Quantum behaviour then emerges when the diffusion strength attains the ``quantum'' value (and classical behaviour in the small-diffusion limit), so the quantum--classical distinction is not a fundamental split but a continuous change of regime governed by the diffusion parameter. Within this framework, individual events are ontological (a single realized trajectory in each run), while only probabilistic predictions are available at the ensemble level. Moreover, the measurement update need not be postulated as an independent projection: one may treat ``collapse'' as a consequence of conditioning on the obtained outcome together with a time-symmetric minimal-change principle selecting the post-measurement diffusion \cite{Pavon1999}. In this sense, irreducible randomness is not merely epistemic but is built into the fundamental description. Further, the usual measurement problem---and much of the subsequent proliferation of ``interpretations'' aimed at reconciling unitary quantum dynamics with definite outcomes---is largely obviated, since outcome events are taken as ontological at the trajectory level while the Schr\"odinger/Hilbert-space formalism functions as an emergent ensemble description.

\medskip
\noindent
\textbf{Role of extended Wigner's Friend.}
Extended Wigner's Friend scenarios serve as a stress test because they require ``friends'' to be treated both as
(i) coherent quantum systems (so that an outside observer assigns a superposed laboratory state) and
(ii) agents with stable records and reasoning capacity.
Any option must therefore state explicitly where (Q1)--(Q3) are addressed and which physical resources (classical data,
irreversibility, and/or added ontology) are being assumed.

\begin{figure}[t]
\centering
\resizebox{0.95\linewidth}{!}{%
\begin{tikzpicture}[
 scale=0.65,
  >=Latex,
  box/.style={draw, rounded corners, align=center, inner sep=5pt,
              font=\small, text width=0.42\linewidth},
  decision/.style={draw, diamond, aspect=2, align=center, inner sep=2pt,
                   font=\small, text width=0.30\linewidth},
  arrow/.style={->, thick},
  node distance=7mm and 10mm
]

\node[scale=0.65,box, text width=0.62\linewidth] (q)
{Question:\\What sort of world allows Darwinian evolution?};

\node[scale=0.65,box, below=of q, text width=0.62\linewidth] (req)
{Evolution requires:\\(E1) stable records \quad (E2) copying + variation \quad (E3) irreversibility};

\node[scale=0.65,box, below=of req, text width=0.62\linewidth] (cqm)
{CQM constraint:\\Copy/delete are available only for a realized \emph{classical data sector}\\(a preferred basis / record structure)};

\node[scale=0.65,box, below left=of cqm] (adlam)
{Agency constraint (Adlam et al.):\\``purely quantum'' (coherent unitary, no decoherence/collapse, no basis)\\$\Rightarrow$ no copying $\Rightarrow$ no minimal agency\\(stressing record-based processes)};

\node[scale=0.65,box, below right=of cqm] (decoh)
{Decoherence / pointer basis:\\explains basis-selection and robust records\\but does not, by itself, select unique outcomes};

\node[scale=0.65,decision, below=of cqm, yshift=-18mm] (ont)
{Ontology of\\histories/facts?};

\node[scale=0.65,box, below left=of ont] (A)
{A: Unique history realism\\collapse / hidden variables\\$\Rightarrow$ single outcomes};

\node[scale=0.65,box, below right=of ont] (B)
{B: Unitary + decohered multiplicity\\records within each decohered history};

\node[scale=0.65,box, below=of A] (C)
{C: Agent-relative facts\\(relational / QBist family)};

\node[scale=0.65,box, below=of B] (D)
{D: Stochastic foundation\\variable diffusion bridges classical--quantum;\\Pavon update via conditioning + minimal drift change};

\node[scale=0.65,box, below=of C, text width=0.62\linewidth, xshift=0.21\linewidth] (wf)
{Extended Wigner's Friend stress test:\\Friends treated both as coherent labs \emph{and} as agents with records.\\Forces explicit placement of (i) record basis, (ii) outcome selection, (iii) agency resources.};

\draw[arrow] (q) -- (req);
\draw[arrow] (req) -- (cqm);
\draw[arrow] (cqm) -- (ont);

\draw[arrow] (adlam.north) -- ++(0,4mm) -| (cqm.south west);
\draw[arrow] (decoh.north) -- ++(0,4mm) -| (cqm.south east);

\draw[arrow] (ont) -- (A);
\draw[arrow] (ont) -- (B);
\draw[arrow] (A) -- (C);
\draw[arrow] (B) -- (D);

\draw[arrow] (C) -- (wf);
\draw[arrow] (D) -- (wf);

\end{tikzpicture}
}
\caption{Road map of the paper (see text for details).}
\label{fig:roadmap}
\end{figure}
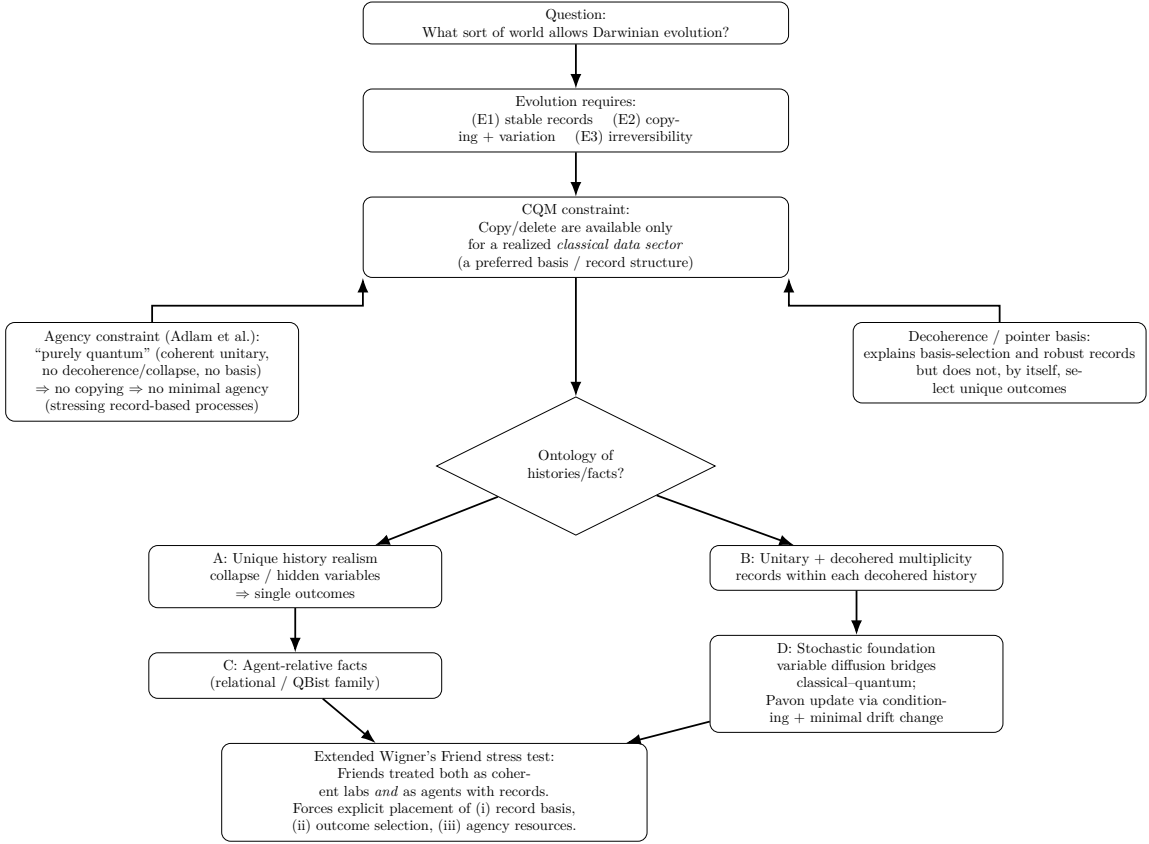

\section{Stochastic mechanics as a foundational option: variable diffusion and conditioning}
Nelson-type stochastic mechanics posits an underlying stochastic trajectory ontology $x(t,\omega)$ with an ensemble description \cite{Nelson1985}. In the standard construction, a particular diffusion strength yields Schr\"odinger dynamics for the ensemble.

A natural unifying idea is to allow the effective diffusion strength (denote it $\nu$) to vary by regime (mass/scale/environment), so that the quantum--classical split becomes a continuum:
\begin{itemize}[itemsep=2pt]
\item $\nu\to 0$: classical Hamilton--Jacobi behaviour (negligible quantum corrections);
\item $\nu\sim \hbar/(2m)$: standard quantum behaviour;
\item intermediate $\nu$: mesoscopic interpolation (in principle testable).
\end{itemize}
On this view, ``classicality'' is not a primitive ontological kind; it is a regime of small effective diffusion.

\medskip
\noindent
\textbf{Collapse as update}:

Pavon \cite{Pavon1999} argues that in the stochastic-mechanics setting, a collapse-like update can be treated as
(i) Bayesian conditioning on the obtained outcome at the ensemble level, followed by
(ii) selection of the post-measurement diffusion that is minimally invasive relative to the pre-measurement dynamics in a
time-symmetric variational sense. This yields a post-measurement wavefunction $\tilde\psi$ (and associated process) consistent with standard reduction rules, without positing collapse as an independent primitive axiom.

\section{Implications for extended Wigner's Friend (summary)}
Extended Wigner's Friend scenarios highlight a tension between two descriptions of an embedded observer:
\begin{enumerate}[label=(W\arabic*),itemsep=2pt]
\item treating the ``friend+lab'' as a coherent quantum system (so that an outside observer assigns a superposed lab state), and
\item treating the friend as an agent possessing stable records and reasoning capacity.
\end{enumerate}
The agency constraint sharpens this tension by emphasizing that strictly coherent ``purely quantum'' regimes lack the copying resources needed for stable record-keeping and deliberation. Accordingly, any proposed ontology must make explicit where the record sector comes from, how (or whether) outcomes are selected, and what physical resources underwrite the use of ``facts'' by agents.

\section{Summary and Conclusions}
Darwinian evolution, at its most basic, presupposes stable records, the capacity to copy those records (with occasional variation), and a reliable arrow of irreversibility through which alternatives are discarded and selection can accumulate. Categorical quantum mechanics helps sharpen this point by making precise why ``copy'' and ``delete'' are not basis-free quantum operations: they are available only relative to a realized \emph{classical-data} sector, i.e.\ a preferred record basis. Decoherence, in turn, explains how such a pointer basis can be selected dynamically and why records become robust, but it does not by itself settle the further question of whether a single definite outcome is selected in each run.

For this reason, it is useful to separate the \emph{record question} from the \emph{outcome question}, and to present in a neutral way the main ontological packages that locate facts and histories differently. The ``agency constraint'' of Adlam--McQueen--Waegell may be read as reinforcing the same structural moral: whenever observers are treated as agents with stable memories and communicable records, substantial classical resources are already being assumed. Extended Wigner's Friend scenarios bring these assumptions into especially sharp focus, since they simultaneously invite a fully coherent quantum description of an entire laboratory and yet require the embedded ``friend'' to function as a record-bearing agent.

Finally, a stochastic-mechanics foundation (in the spirit of Nelson) offers a distinct way to organise these issues: classical stochastic trajectories are taken as fundamental, quantum behaviour arises at the ensemble level, and measurement update can be treated not as an added postulate but as conditioning supplemented by a principled minimal-change rule. On this view the quantum--classical divide is approached as a continuum of regimes (e.g.\ via variable diffusion), while retaining an ontology of individual events together with inherently probabilistic predictions.

\section*{Acknowledgements}
I acknowledge the use of ChatGPT (OpenAI) as a language and structuring aid. All concepts, arguments, and claims in this paper--and any remaining errors--are solely my responsibility.


\begin{thebibliography}{99}

\bibitem{WZ1982}
W.~K.~Wootters and W.~H.~Zurek,
``A single quantum cannot be cloned,''
\emph{Nature} \textbf{299} (1982) 802--803.

\bibitem{PBNoDelete2000}
A.~K.~Pati and S.~L.~Braunstein,
\newblock ``Impossibility of deleting an unknown quantum state,''
\newblock \emph{Nature} \textbf{404} (6774) (2000) 164--165.
\newblock DOI: \href{https://doi.org/10.1038/404130b0}{10.1038/404130b0}.
\newblock Preprint: \href{https://arxiv.org/abs/quant-ph/9911090}{arXiv:quant-ph/9911090}.


\bibitem{CK2017}
B.~Coecke and A.~Kissinger,
\emph{Picturing Quantum Processes},
Cambridge University Press (2017).

\bibitem{Zurek2003}
W.~H.~Zurek,
``Decoherence, einselection, and the quantum origins of the classical,''
\emph{Rev.\ Mod.\ Phys.} \textbf{75} (2003) 715--775.

\bibitem{FR2018}
D.~Frauchiger and R.~Renner,
\newblock ``Quantum theory cannot consistently describe the use of itself,''
\newblock \emph{Nature Communications} \textbf{9} (2018) 3711.
\newblock DOI: \href{https://doi.org/10.1038/s41467-018-05739-8}{10.1038/s41467-018-05739-8}.
\newblock Preprint: \href{https://arxiv.org/abs/1604.07422}{arXiv:1604.07422}.

\bibitem{Brukner2018}
\v{C}.~Brukner,
\newblock ``A no-go theorem for observer-independent facts,''
\newblock \emph{Entropy} \textbf{20} (5) (2018) 350.
\newblock DOI: \href{https://doi.org/10.3390/e20050350}{10.3390/e20050350}.
\newblock Preprint: \href{https://arxiv.org/abs/1804.00749}{arXiv:1804.00749}.

\bibitem{Proietti2019}
M.~Proietti, A.~Pickston, F.~Graffitti, P.~Barrow, D.~Kundys, C.~Branciard,
M.~Ringbauer, and A.~Fedrizzi,
\newblock ``Experimental test of local observer independence,''
\newblock \emph{Science Advances} \textbf{5} (9) (2019) eaaw9832.
\newblock DOI: \href{https://doi.org/10.1126/sciadv.aaw9832}{10.1126/sciadv.aaw9832}.
\newblock Preprint: \href{https://arxiv.org/abs/1902.05080}{arXiv:1902.05080}.


\bibitem{Adlam2025}
E.~C.~Adlam, K.~J.~McQueen, and M.~Waegell,
``Agency cannot be a purely quantum phenomenon,''
arXiv:2510.13247 (2025).
 

\bibitem{Nelson1985}
E.~Nelson,
\emph{Quantum Fluctuations},
Princeton University Press (1985).

\bibitem{Pavon1999}
M.~Pavon,
``Derivation of the wave function collapse in the context of Nelson's stochastic mechanics,'' \emph{J. Math. Physics}, \textbf{40}
 (1999), 5565-5577.
\newblock DOI: https://doi.org/10.1063/1.533046.

   
\bibitem{GRW1986}
G.~C.~Ghirardi, A.~Rimini, and T.~Weber,
\newblock ``Unified dynamics for microscopic and macroscopic systems,''
\newblock \emph{Physical Review D} \textbf{34} (2) (1986) 470--491.
\newblock DOI: \href{https://doi.org/10.1103/PhysRevD.34.470}{10.1103/PhysRevD.34.470}.

\bibitem{Pearle1989}
P.~Pearle,
\newblock ``Combining stochastic dynamical state-vector reduction with spontaneous localization,''
\newblock \emph{Physical Review A} \textbf{39} (5) (1989) 2277--2289.
\newblock DOI: \href{https://doi.org/10.1103/PhysRevA.39.2277}{10.1103/PhysRevA.39.2277}.

\bibitem{GPR1990}
G.~C.~Ghirardi, P.~Pearle, and A.~Rimini,
\newblock ``Markov processes in Hilbert space and continuous spontaneous localization of systems of identical particles,''
\newblock \emph{Physical Review A} \textbf{42} (1) (1990) 78--89.
\newblock DOI: \href{https://doi.org/10.1103/PhysRevA.42.78}{10.1103/PhysRevA.42.78}.

\bibitem{Diosi1989}
L.~Di\'osi,
\newblock ``Models for universal reduction of macroscopic quantum fluctuations,''
\newblock \emph{Physical Review A} \textbf{40} (3) (1989) 1165--1174.
\newblock DOI: \href{https://doi.org/10.1103/PhysRevA.40.1165}{10.1103/PhysRevA.40.1165}.

\bibitem{Penrose1996}
R.~Penrose,
\newblock ``On Gravity's role in Quantum State Reduction,''
\newblock \emph{General Relativity and Gravitation} \textbf{28} (5) (1996) 581--600.
\newblock DOI: \href{https://doi.org/10.1007/BF02105068}{10.1007/BF02105068}.

\bibitem{BassiGhirardi2003}
A.~Bassi and G.~C.~Ghirardi,
\newblock ``Dynamical reduction models,''
\newblock \emph{Physics Reports} \textbf{379} (5--6) (2003) 257--426.
\newblock DOI: \href{https://doi.org/10.1016/S0370-1573(03)00103-0}{10.1016/S0370-1573(03)00103-0}.

\bibitem{BassiEtal2013}
A.~Bassi, K.~Lochan, S.~Satin, T.~P.~Singh, and H.~Ulbricht,
\newblock ``Models of wave-function collapse, underlying theories, and experimental tests,''
\newblock \emph{Reviews of Modern Physics} \textbf{85} (2013) 471--527.
\newblock DOI: \href{https://doi.org/10.1103/RevModPhys.85.471}{10.1103/RevModPhys.85.471}.
\newblock Preprint: \href{https://arxiv.org/abs/1204.4325}{arXiv:1204.4325}.

\bibitem{Bohm1952a}
D.~Bohm,
\newblock ``A Suggested Interpretation of the Quantum Theory in Terms of ``Hidden'' Variables. I,''
\newblock \emph{Physical Review} \textbf{85} (1952) 166--179.
\newblock DOI: \href{https://doi.org/10.1103/PhysRev.85.166}{10.1103/PhysRev.85.166}.

\bibitem{Bohm1952b}
D.~Bohm,
\newblock ``A Suggested Interpretation of the Quantum Theory in Terms of ``Hidden'' Variables. II,''
\newblock \emph{Physical Review} \textbf{85} (1952) 180--193.
\newblock DOI: \href{https://doi.org/10.1103/PhysRev.85.180}{10.1103/PhysRev.85.180}.

\bibitem{DGZ2013}
D.~D\"urr, S.~Goldstein, and N.~Zangh\`{\i},
\newblock ``Quantum Physics Without Quantum Philosophy,''
\newblock \emph{Springer}, Berlin, 2013.
\newblock DOI: \href{https://doi.org/10.1007/978-3-642-30690-7}{10.1007/978-3-642-30690-7}.

\bibitem{Holland}
P.~ R.~ Holland, \emph{The Quantum Theory of Motion}, Cambridge, (1993). 

\bibitem{Griffiths1984}
R.~B.~Griffiths,
\newblock ``Consistent histories and the interpretation of quantum mechanics,''
\newblock \emph{Journal of Statistical Physics} \textbf{36} (1--2) (1984) 219--272.
\newblock DOI: https://doi.org/10.1007/BF01015734.

\bibitem{GMH1990}
M.~Gell-Mann and J.~B.~Hartle,
\newblock ``Quantum mechanics in the light of quantum cosmology,''
\newblock in W.~H.~Zurek (ed.), \emph{Complexity, Entropy, and the Physics of Information}
(Santa Fe Institute Studies in the Sciences of Complexity, Vol.~VIII),
Addison--Wesley, Reading, MA, 1990, pp.~425--458.

\bibitem{GMH1993}
M.~Gell-Mann and J.~B.~Hartle,
\newblock ``Classical equations for quantum systems,''
\newblock \emph{Physical Review D} \textbf{47} (8) (1993) 3345--3382.
\newblock DOI: \href{https://doi.org/10.1103/PhysRevD.47.3345}{10.1103/PhysRevD.47.3345}.
\newblock Preprint: \href{https://arxiv.org/abs/gr-qc/9210010}{arXiv:gr-qc/9210010}.

\bibitem{Rovelli1996}
C.~Rovelli,
\newblock ``Relational quantum mechanics,''
\newblock \emph{International Journal of Theoretical Physics} \textbf{35} (8) (1996) 1637--1678.
\newblock DOI: \href{https://doi.org/10.1007/BF02302261}{10.1007/BF02302261}.
\newblock Preprint: \href{https://arxiv.org/abs/quant-ph/9609002}{arXiv:quant-ph/9609002}.

\bibitem{CFS2002}
C.~M.~Caves, C.~A.~Fuchs, and R.~Schack,
\newblock ``Quantum probabilities as Bayesian probabilities,''
\newblock \emph{Physical Review A} \textbf{65} (2002) 022305.
\newblock DOI: \href{https://doi.org/10.1103/PhysRevA.65.022305}{10.1103/PhysRevA.65.022305}.
\newblock Preprint: \href{https://arxiv.org/abs/quant-ph/0106133}{arXiv:quant-ph/0106133}.

\bibitem{Fuchs2010}
C.~A.~Fuchs,
\newblock ``QBism, the Perimeter of Quantum Bayesianism,''
\newblock Preprint: \href{https://arxiv.org/abs/1003.5209}{arXiv:1003.5209}.

\bibitem{FS2013}
C.~A.~Fuchs and R.~Schack,
\newblock ``Quantum-Bayesian coherence,''
\newblock \emph{Reviews of Modern Physics} \textbf{85} (2013) 1693--1715.
\newblock DOI: \href{https://doi.org/10.1103/RevModPhys.85.1693}{10.1103/RevModPhys.85.1693}.
\newblock Preprint: \href{https://arxiv.org/abs/1301.3274}{arXiv:1301.3274}.

\bibitem{FMS2014}
C.~A.~Fuchs, N.~D.~Mermin, and R.~Schack,
\newblock ``An introduction to QBism with an application to the locality of quantum mechanics,''
\newblock \emph{American Journal of Physics} \textbf{82} (8) (2014) 749--754.
\newblock DOI: \href{https://doi.org/10.1119/1.4874855}{10.1119/1.4874855}.
\newblock Preprint: \href{https://arxiv.org/abs/1311.5253}{arXiv:1311.5253}.


\end{thebibliography}
\end{document}